\documentclass[%
 reprint,
nofootinbib,
amsmath,amssymb,
 aps,
pra,
]{revtex4-2}
\usepackage{amsmath}
\usepackage{xcolor}
\usepackage{hyperref}
\usepackage{titlesec}
\usepackage{caption}
\usepackage{subcaption}
\usepackage{graphicx}
\usepackage{ragged2e} % for \justifying
\usepackage[justification=justified,singlelinecheck=false]{caption}

\usepackage[export]{adjustbox}
\usepackage{flushend}
\usepackage{booktabs}
\usepackage{tabularx}
\usepackage{multirow}
\usepackage{cleveref}
\usepackage{dcolumn}% Align table columns on decimal point
\usepackage{bm}% bold math
\usepackage{algorithm2e}
\usepackage{algorithmic}%
\usepackage{amssymb}
\usepackage{ulem}
\RestyleAlgo{ruled}
\usepackage{caption}
\usepackage{color}
\usepackage{appendix}

\begin{document}
% Tighter spacing around displayed equations
\setlength\abovedisplayskip{6pt}
\setlength\belowdisplayskip{6pt}
\setlength\abovedisplayshortskip{4pt}
\setlength\belowdisplayshortskip{4pt}

%\preprint{APS/123-QED}

\title{Force Sensing Beyond the Standard Quantum Limit in a Hybrid
Optomechanical Platform}% Force line breaks with \\
\author{Alolika Roy , Amarendra K. Sarma }
 %   \email[Correspondence email address]
{ \affiliation{ 
     Department of Physics, Indian Institute of Technology Guwahati, Guwahati 781039, India.}
% \author{Alolika Roy $^{1,2}$, Amarendra K. Sarma $^1$}
%  %   \email[Correspondence email address]{anksing@iitk.ac.in}
% { \affiliation{ 
%     $^1$ Department of Physics, Indian Institute of Technology Guwahati, Guwahati 781039, India.\\
%     $^2$ Department of Physics, B.D College, Patna.}
\date{\today}% It is always \today, today,
             %  but any date may be explicitly speci fied
\begin{abstract}
We theoretically investigate quantum measurement noise in a hybrid optomechanical system, focusing on radiation pressure back action and its impact on force sensing. The setup consists of an optomechanical cavity with a movable mirror, a fixed semi transparent mirror, an ensemble of quantum dots (QD) coupled to the cavity mode, and an intracavity optical parametric amplifier (OPA). We show how the QD induced response, together with the system nonlinearity, modifies the noise spectral density and thereby improves the force measurement sensitivity. In this setup, coherent quantum noise cancellation (CQNC) can completely remove the back action noise. In addition, increasing the OPA pump gain enables sensitivity beyond the standard quantum limit (SQL) at reduced laser power. These combined effects allow weak force sensing beyond the SQL.

\end{abstract}

\maketitle

\section{Introduction}
Quantum interference and coherence underpin a range of striking quantum-optical phenomena, including coherent population trapping, electromagnetically induced transparency, light storage, and quantum noise suppression\citep{Gray:78,PhysRevLett.66.2593,Liu2001,bayati2025optomechanically, PhysRevD.23.1693,RevModPhys.77.633,Zhu:20}. These effects not only deepen our understanding of fundamental light–matter interactions but also provide powerful tools for manipulating quantum systems in practical applications~\citep{RevModPhys.82.1041,RevModPhys.75.457}.  In particular, they offer routes to engineer measurement dynamics and quantum noise in precision metrology, including cavity-optomechanical force sensing~\citep{RevModPhys.86.1391,LiOuLeiLiu+2021+2799+2832,10.3389/frqst.2022.1091691,tsang2010coherent}. A major challenge in this context arises from measurement of back-action noise, which fundamentally limits the precision of weak-force detection \citep{RevModPhys.82.1155,PhysRevD.23.1693,harry2010advanced,RevModPhys.52.341}. In cavity optomechanical systems, quantum fluctuations of the optical field impart radiation-pressure noise onto the mechanical oscillator, setting the standard quantum limit (SQL) \citep{bowen2015quantum,caves1980measurement,clerk2010introduction,danilishin2012quantum,RevModPhys.86.1391} and restricting the attainable measurement sensitivity \citep{caves1980measurement,danilishin2012quantum,RevModPhys.86.1391}.

Recent progress in the engineering of solid-state quantum systems particularly ensembles of quantum dots (QDs) embedded in high-finesse optical cavities \citep{Reithmaier2004,yadav2023optical,Heindel:23,RevModPhys.87.347,https://doi.org/10.1002/qute.202300390,Sarma:16} has opened new avenues for controlling quantum noise and realizing enhanced sensing protocols\citep{s7123489,karnieli2023quantum,al2025quantum}. Experiments have demonstrated robust optical coupling of quantum dots to high-finesse cavities, including Purcell-enhanced emission from QD ensembles in microcavities \citep{PhysRevB.78.235306}. Strong light–matter coupling has also been observed in photonic-crystal cavities with multiple QDs and in high-$Q$ microcavities containing colloidal QDs \citep{Kim:11,10.1063/1.3558731}. The strong coupling between cavity photons and excitonic states in QD ensembles substantially modifies the cavity response \citep{PhysRevLett.101.083601,dovzhenko2021strong}, creating additional interference channels that can be exploited to manipulate noise and dissipation. Such hybrid cavity–QD systems~\citep{Motazedifard_2016,bariani2015atom,singh2023enhanced,Gupta:24,roy2025overcoming,Singh:14,Li:19} enable precise control over optical nonlinearities, energy decay pathways, and quantum correlations, thereby offering a flexible and scalable platform for coherent control of noise processes\citep{RevModPhys.87.347,s7123489,karnieli2023quantum,QU2025130910}.

A particularly promising approach to surpass the SQL \citep{bondurant1986reduction} is coherent quantum noise cancellation (CQNC)\citep{tsang2010coherent}, which relies on destructive interference between intrinsic radiation-pressure noise and auxiliary noise channels engineered within the system \citep{Motazedifard_2016,bariani2015atom,singh2023enhanced}. When implemented in hybrid optomechanical architectures incorporating atomic ensembles\citep{Motazedifard_2016,bariani2015atom,singh2023enhanced,Bemani:15} and superconducting qubit \citep{nongthombam2023synchronization,PhysRevA.104.023509,roy2025overcoming}, CQNC can be achieved by tailoring the photon–exciton–phonon interactions to cancel the dominant back-action noise. This mechanism enables the mechanical oscillator to detect weak external forces with sensitivities beyond the conventional quantum limit~\citep{PhysRevA.73.033819,Shomroni2019}.Noise reduction in hybrid systems has also been demonstrated experimentally
 \cite{moller2017quantum}.

Motivated by these advances, we theoretically investigate weak-force sensing in a hybrid optomechanical cavity coupled to an ensemble of quantum dots. We show that an engineered interference pathway enables coherent quantum noise cancellation, leading to complete suppression of measurement back action and force sensitivities beyond the standard quantum limit. We employ the quantum Langevin formalism and solve the linearized system dynamics. Using these solutions, we evaluate the output noise spectra and identify the operating regime where full radiation pressure noise cancellation is achieved by tuning the cavity detuning and the effective exciton mechanical coupling. We further find that an intracavity Optical Parametric Amplifier (OPA) ~\citep{peano2015intracavity, singh2023enhanced} enhances the low frequency response, enabling high sensitivity at reduced drive power. Together, these results establish this hybrid platform as a practical route to SQL beating performance for precision metrology and weak force detection.

% Motivated by these advances, we theoretically study how coupling an optomechanical cavity to an ensemble of quantum dots can enhance force sensitivity, with particular emphasis on implementing CQNC to suppress back-action noise. Using the quantum Langevin formalism, we analyze the system dynamics and noise spectra, it is demonstrating that appropriate tuning of the cavity detuning and exciton–mechanical coupling parameters can lead to complete radiation-pressure noise cancellation. Our results reveal that such hybrid optomechanical–semiconductor systems offer a robust and experimentally feasible route towards SQL-beating performance, establishing them as promising candidates for next-generation quantum sensing, precision metrology, and weak-force detection beyond the reach of conventional back-action–limited platforms. 
\section{System}
\label{system}
\begin{figure}[h!]
\centering
    \includegraphics[height=5cm,clip]{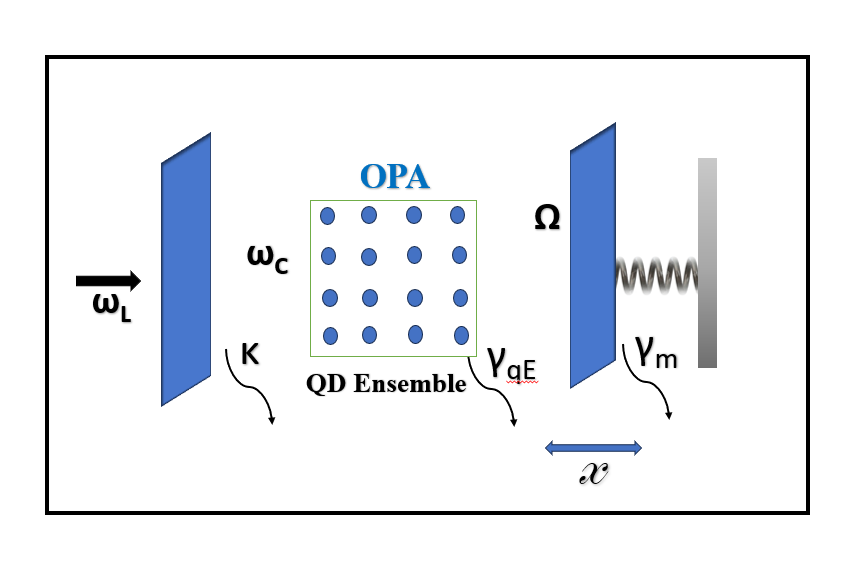}
\captionsetup{justification=raggedright}
\caption{Schematic diagram of a Hybrid optomechanical system equipped with an ensemble of quantum dots and Optical Parametric Amplifier (OPA). }
\end{figure}
This work analyzes a hybrid optomechanical system consisting of an optical cavity with resonance frequency $\omega_{c}$, a mechanical oscillator of frequency $\Omega$, and an ensemble of quantum dots (QDs) coupled to the cavity mode. The system is engineered such that the QD ensemble interacts only with the optical field and remains decoupled from the mechanical oscillator. The optomechanical cavity is coherently driven by a laser with frequency $\omega_{L}$ and input power $P_{L}$. A schematic representation of the full system configuration is shown in Fig.~1.

The total Hamiltonian of the system can be written as
\begin{equation}
H = H_{\text{OM}} + H_{\text{OPA}} + H_{\text{L}} + H_{\text{QD}},
\label{eq:Htotal}
\end{equation}
where each term represents a different subsystem or interaction within the hybrid platform.

The optomechanical Hamiltonian is given by
\begin{equation}
H_{\text{OM}} = \hbar \Omega \hat{b}^\dagger \hat{b} + \hbar \Delta_c \hat{a}^\dagger \hat{a}
- \hbar g_0 \hat{a}^\dagger \hat{a}(\hat{b}^\dagger + \hat{b}),
\end{equation}
where $\Omega$ is the frequency of the mechanical oscillator. This Hamiltonian describes the standard interaction between the mechanical and optical modes. The first term corresponds to the mechanical mode energy, the second to the optical cavity field, and the third term represents their radiation-pressure coupling with strength $g_0$. Here, the detuning between the laser and cavity frequencies is $\Delta_c = \omega_L - \omega_c$. The operators $\hat{a}$ ($\hat{a}^\dagger$) and $\hat{b}$ ($\hat{b}^\dagger$) are the annihilation (creation) operators for the cavity and mechanical modes, respectively.

The contribution of the optical parametric amplifier (OPA) is expressed as,
\begin{equation}
H_{\text{OPA}} = i\hbar \mathcal{G}\,(\hat{a}^{\dagger 2} e^{i\theta} - \hat{a}^2 e^{-i\theta}),
\end{equation}
where $\mathcal{G}$ is the gain of the OPA and $\theta$ denotes the pump phase. Setting $\theta = 0$ simplifies the term to
\begin{equation}
H_{\text{OPA}} = i\hbar \mathcal{G}\,(\hat{a}^{\dagger 2} - \hat{a}^2).
\end{equation}

The Hamiltonian describing the laser drive is given by,
\begin{equation}
H_{\text{L}} = i\hbar E_L (\hat{a}^\dagger - \hat{a}),
\end{equation}
where $E_L$ represents the amplitude of the external driving field applied to the optical cavity.

Finally, the quantum-dot ensemble Hamiltonian can be expressed as (see Appendix \ref{app:QDensemble} for details)
\begin{equation}
H_{\text{QD}} = \hbar \omega_{qe}\, \hat{S}^\dagger \hat{S} + \frac{1}{2}\hbar G_{cs}\,(\hat{a}^\dagger + \hat{a})(\hat{S}^\dagger + \hat{S}),
\end{equation}
where $\hat{S}$ ($\hat{S}^\dagger$) are the collective annihilation (creation) operators representing the excitation mode of the quantum-dot ensemble. 
The first term describes the free energy of the QD ensemble with transition frequency $\omega_{qe}$, while the second term captures the dipole interaction between the optical field and the collective QD mode, characterized by the coupling strength $G_{cs}$.

Moving to the frame rotating at the drive frequency $\omega_L$
and dropping the constants, the total Hamiltonian becomes
\begin{equation}
\begin{aligned}
H &= -\hbar \Delta_c\, \hat a^\dagger \hat a
    + \hbar \Omega\, \hat b^\dagger \hat b
    - \hbar \Delta_{qe}\, \hat S^\dagger \hat S
    + i\hbar \mathcal{G}\,(\hat{a}^{\dagger 2} e^{i\theta} - \hat{a}^2 e^{-i\theta})\\
  &\quad + \hbar g_0 \hat a^\dagger \hat a\,(\hat b+\hat b^\dagger)
    + \frac{\hbar G_{cs}}{2}\,(\hat a+\hat a^\dagger)(\hat S+\hat S^\dagger)\\
    &\quad + i\hbar E_L(\hat a^\dagger - \hat a),
\end{aligned}
\label{eq:Hrot}
\end{equation}
with detunings $\Delta_c = \omega_L - \omega_c$ and $\Delta_{qe} = \omega_L - \omega_{qe}$.

% In the low-excitation regime, the collective excitations of the QD ensemble behave as bosonic quasi-particles, obeying the commutation relation $[\hat{S}, \hat{S}^\dagger] = 1$.

In the linearized description, each system operator is expressed as a steady-state mean value plus a small quantum fluctuation,
$\hat{A}(t) = \bar{A} + \delta \hat{A}(t)$. Terms higher than first order in the fluctuations can be neglected. The amplitude and phase quadratures of the optical, mechanical, and qubit modes can be defined as
$
    \hat{x}_A = \frac{\hat{A} + \hat{A}^\dagger}{\sqrt{2}}, \quad
    \hat{p}_A = i\,\frac{\hat{A}^\dagger - \hat{A}}{\sqrt{2}},
$
with $\hat{A} = \hat{a}, \hat{b}, \hat{S}$ and $\bar{A} = \alpha, \beta, {\alpha_S}$. The corresponding input noise operators are denoted by 
$\hat{A}^{\mathrm{in}} = \hat{a}^{\mathrm{in}}, \hat{b}^{\mathrm{in}}, \hat{S}^{\mathrm{in}}$  and the amplitude and phase of noise terms are introduced as
$
    \hat{x}^{\mathrm{in}}_A = \frac{\hat{A}^{\mathrm{in}} + \hat{A}^{\mathrm{in}\dagger}}{\sqrt{2}},
    \qquad
    \hat{p}^{\mathrm{in}}_A = i\,\frac{\hat{A}^{\mathrm{in}\dagger} - \hat{A}^{\mathrm{in}}}{\sqrt{2}}.
$
They satisfy the usual Markovian correlations~\cite{walls1994gj,benguria1981quantum,gardiner1985input,wimmer2014coherent} as follows
$$
    \langle \hat{a}_{\mathrm{in}}(t)\,\hat{a}^\dagger_{\mathrm{in}}(t') \rangle
    =
    \langle \hat{b}_{\mathrm{in}}(t)\,\hat{b}^\dagger_{\mathrm{in}}(t') \rangle
    =
    \langle \hat{S}_{\mathrm{in}}(t)\,\hat{S}^\dagger_{\mathrm{in}}(t') \rangle
    =
    \delta(t - t').
$$
In terms of these fluctuation and noise quadratures, the linearized quantum Langevin equations can be written as
\begin{align}
\dot{\hat{x}}_b &= \Omega \hat{p}_b, \\
\dot{\hat{p}}_b &= -\Omega \hat{x}_b - \gamma_m \hat{p}_b - g \hat{x}_a + \sqrt{2\gamma_m}\,(\hat{F}_{\mathrm{th}} + F_{\mathrm{ext}}), \\
\dot{\hat{x}}_a &= \left(-\frac{\kappa}{2} + 2\mathcal{G}\right)\hat{x}_a + \sqrt{\kappa}\,\hat{x}_{a,\mathrm{in}}, \\
\dot{\hat{p}}_a &= -g \hat{x}_b + \left(-\frac{\kappa}{2} - 2\mathcal{G}\right)\hat{p}_a - G_{cs}\hat{x}_S + \sqrt{\kappa}\,\hat{p}_{a,\mathrm{in}}, \\
\dot{\hat{x}}_S &= -\frac{\gamma_{qE}}{2}\hat{x}_S - \Delta_{qe} \hat{p}_S + \sqrt{\gamma_{qE}}\,\hat{x}_{S,\mathrm{in}}, \\
\dot{\hat{p}}_S &= -G_{cs}\hat{x}_a + \Delta_{qe} \hat{x}_S - \frac{\gamma_{qE}}{2}\hat{p}_S + \sqrt{\gamma_{qE}}\,\hat{p}_{S,\mathrm{in}}.
\end{align}

Here, $g =\sqrt{2} g_0 \alpha$, with $\alpha $ denoting the steady-state mean value of the cavity field operator $\hat{a}$. $f_{\mathrm{th}}$ represents the Brownian thermal noise and $f_{\mathrm{ext}}$ denotes the external force acting on the mechanical oscillator. 
The mechanical damping rate, cavity decay rate, and dephasing rate of the QD ensemble are given by $\gamma_m$, $\kappa$, and $\gamma_{qE}$, respectively.

The linear response of the system is obtained by moving from the time domain to the frequency domain \cite{RevModPhys.86.1391}. Using the Fourier transform
$ \hat{X}(\omega) = \frac{1}{\sqrt{2\pi}} \int \! dt \,  \hat{x}(t)\, e^{-i\omega t}$, the Langevin equations in the Fourier domain take the following form

\begin{align}
\hat X_b(\omega) &= \chi_m(\omega)\!\left[-\,g\,\hat X_a(\omega)
+ \sqrt{2\gamma_m}\,\big(\hat F_{\rm th}(\omega)+F_{\rm ext}(\omega)\big)\right], \\
\hat P_b(\omega) &= \frac{i\omega}{\Omega}\,\hat X_b(\omega),\\
\hat X_a(\omega) &= \sqrt{\kappa}\,\lambda_+(\omega)\,\hat X_{a,{\rm in}}(\omega), \\
\hat P_a(\omega) &= g^2 \chi_m(\omega)\lambda_-(\omega)\lambda_+(\omega)\sqrt{\kappa}\,\hat X_{a,{\rm in}}(\omega)\notag\\
&- g\,\chi_m(\omega)\lambda_-(\omega)\sqrt{2\gamma_m}\,\big(\hat F_{\rm th}(\omega)+F_{\rm ext}(\omega)\big) \notag \\ & -G_{cs}\lambda_-(\omega)\,\hat X_S(\omega)
+ \lambda_-(\omega)\sqrt{\kappa}\,\hat P_{a,{\rm in}}(\omega), \\
\hat X_S(\omega) &= -\,\Delta_{qe}\,\chi_S(\omega)\,\hat P_S(\omega)
+ \chi_S(\omega)\sqrt{\gamma_{qE}}\,\hat X_{S,{\rm in}}(\omega), \\
\hat P_S(\omega) &= -\,G_{cs}\,\xi(\omega)\lambda_+(\omega)\sqrt{\kappa}\,\hat X_{a,{\rm in}}(\omega)
- \notag \\
&-
\chi'_S(\omega)\sqrt{\gamma_{qE}}\,\hat X_{S,{\rm in}}(\omega)+
\xi(\omega)\sqrt{\gamma_{qE}}\,\hat P_{S,{\rm in}}(\omega). 
\end{align}
 where, 
 \begin{align*}
&\chi_m(\omega)=\frac{\Omega}{\Omega^2-\omega^2+i\gamma_m\omega},\qquad
\chi_a(\omega)=\frac{1}{i\omega+\kappa/2},\\
&\chi_S(\omega)=\frac{1}{i\omega+\gamma_{qE}/2},
\quad
\xi(\omega)=\Big(i\omega+\frac{\gamma_{qE}}{2}+\Delta_{qe}^{\,2}\,\chi_S\Big)^{-1},\\
&\chi'_S(\omega)=-\,\Delta_{qe}\,\xi(\omega)\,\chi_S(\omega),
\lambda_\pm(\omega)=\Big(\chi_a^{-1}(\omega)\mp 2\mathcal{G}\Big)^{-1},
\quad\\
&g=\sqrt{2}\,g_0\,\alpha, \qquad
G_{cs}'=\sqrt{2}\,G_{cs}\,\alpha_s.
\end{align*}

Here, \(\Delta_{qe}=\omega_L-\omega_{qe}\) is the drive–QD detuning, \(\gamma_{qE}\) is the collective QD linewidth, and \(\alpha\) (\(\alpha_s\)) are the steady-state cavity (QD) amplitudes obtained from the linearized steady-state equations.
The external and thermal forces are conveniently expressed in dimensionless form as
$
    F_{\mathrm{ext}} = \frac{f_{\mathrm{ext}}}{\sqrt{\hbar m \Omega \gamma_m}},
    \quad
    F_{\mathrm{th}}  = \frac{f_{\mathrm{th}}}{\sqrt{\hbar m \Omega \gamma_m}},
$
where $f_{\mathrm{ext}}$ and $f_{\mathrm{th}}$ denote the physical external and thermal forces, respectively~\cite{allahverdi2022homodyne}. The rescaled thermal Brownian force obeys the correlation
$
    \langle F_{\mathrm{th}}(t) F_{\mathrm{th}}(t') \rangle
    = \bar{n}\,\delta(t - t'),
$
with
$
    \bar{n} = \frac{K_B T}{\hbar \Omega},
$
being the mean thermal phonon number of the mechanical mode~\cite{wimmer2014coherent}.

% Scaled force and thermal noise (unchanged in form),
% $F_{\rm ext}(\omega)=\frac{f_{\rm ext}(\omega)}{\sqrt{\hbar m\Omega \gamma_m}}  \rm {and} 
% F_{\rm th}(\omega)=\frac{f_{\rm th}(\omega)}{\sqrt{\hbar m\,\Omega\,\gamma_m}},
% $
% with Brownian force correlations (in the high-$T$/Markov limit used for spectra) 
% \(\langle F_{\rm th}(t)F_{\rm th}(t')\rangle=\bar n\,\delta(t-t')\),
% \(\bar n=\frac{k_B T}{\hbar\Omega}\).

\section{Force sensing with coherent quantum noise cancellation}
\label{CQNC}
When the external force $F_{\mathrm{ext}}$ acts on the movable mirror, it displaces the mechanical oscillator and thereby changes the cavity length. This modulation of the cavity length imprints a phase shift on the intracavity and output optical fields. To quantify this effect, we solve the linearized Langevin equations in the frequency domain and obtain the phase quadrature of the intracavity field, $P_a(\omega)$ (see Appendix \ref{appendixB}). The corresponding output phase quadrature is then given by the standard input--output relation
\begin{equation}
    P_a^{\mathrm{out}}(\omega) = \sqrt{\kappa}\,P_a(\omega) - P_a^{\mathrm{in}}(\omega),
\end{equation}
where $\kappa$ is the cavity decay rate and $P_a^{\mathrm{in}}(\omega)$ denotes the input phase quadrature. This can be expressed as,

Then
\begin{align}
P^{\rm out}_a(\omega)&={}-\,g\,\chi_m(\omega)\,\lambda_{-}(\omega)\,\sqrt{2\gamma_m\kappa}\,
\big[\hat F_{\rm th}(\omega)+F_{\rm ext}(\omega)\big]
\nonumber+\\ \notag
&\big[g^2\chi_m(\omega)+G_{cs}^2\chi'_S(\omega)\big]\lambda_{+}(\omega)\lambda_{-}(\omega)\,
\kappa\,\hat x_{a,{\rm in}}(\omega)+\\&\big(\lambda_{-}(\omega)\kappa-1\big)\hat p_{a,{\rm in}}(\omega)
\nonumber-\,G_{cs}\lambda_{-}(\omega)\sqrt{\kappa\gamma_{qE}}\,\\ \notag
&\Big[\chi_S(\omega)\big(\Delta_{qe}\chi'_S(\omega)+1\big)\hat x_{S,{\rm in}}(\omega)\\
&+\chi'_S(\omega)\hat P_{S,{\rm in}}(\omega)\Big].
\end{align}

where, 
\begin{align}
   &\chi ^\prime_S= -\Omega\chi_S\xi = \frac{\Omega}{\Omega ^2 - \omega^2 + i \omega \gamma_{qE} + \gamma_{qE} ^2/4}                            \\
   &\chi_S=  \frac{1}{i\omega+\gamma_{qE}/2}                                    \\
   &\chi_m=\frac{\Omega}{\Omega^2-\omega^2+i\gamma_m\omega}
\end{align}
The second term of $P^{out}_a(\omega)$ corresponds to the back-action noise \cite{allahverdi2022homodyne,Motazedifard_2016}. For significant noise reduction, we aim to eliminate the back-action noise completely \cite{tsang2010coherent,wimmer2014coherent}.
Thus complete back-action cancellation requires
\begin{equation}
g^2 \chi_m(\omega) + G_{cs}^2 \chi'_S(\omega) = 0,
\label{eq:CQNC_condition}
\end{equation}
This is the CQNC condition and it requires that $g = G_{cs}$ and $\chi_m(\omega) = -\chi'_S(\omega)$ (ideal matching).
This implies that: (i) the response of the mechanical oscillator to the back-action noise is exactly equal in magnitude and opposite in sign to that of the QD-ensemble mode in the system, and (ii) the optomechanical coupling is equal to the effective cavity--QD-ensemble coupling of the hybrid system.

Under this condition, $P^{out}_a(\omega)$ becomes,
\begin{align}
 P^{\rm out}_a(\omega)
&={}-\,g\,\chi_m(\omega)\,\lambda_{-}(\omega)\,\sqrt{2\gamma_m\kappa}\,
\big[\hat F_{\rm th}(\omega)+F_{\rm ext}(\omega)\big]
\nonumber\\
&+\big(\lambda_{-}(\omega)\kappa-1\big)\hat p_{a,{\rm in}}(\omega)
\nonumber-\,G_{cs}\lambda_{-}(\omega)\sqrt{\kappa\gamma_{qE}}\,\notag\\
&\Big[\chi_S(\omega)\big(\Delta_{qe}\chi'_S(\omega)+1\big)\hat x_{S,{\rm in}}(\omega) \notag\\
&+\chi'_S(\omega)\hat P_{S,{\rm in}}(\omega)\Big].
\end{align}

With re-arranging the terms we can write this as follows
\begin{equation}
F_{\rm ext}(\omega)+\hat F_{\rm add}(\omega)
=\frac{-1}{g\,\chi_m(\omega)\,\lambda_{-}(\omega)\,\sqrt{2\gamma_m\kappa}}\;
\hat p^{\rm out}_a(\omega),
\end{equation}
with the added force as
\begin{align}
\label{faddd}
&\hat F_{\rm add}(\omega)
={}\hat F_{\rm th}(\omega)
-\frac{\lambda_{-}(\omega)\kappa-1}{g\,\chi_m(\omega)\,\lambda_{-}(\omega)\,\sqrt{2\gamma_m\kappa}}\,
\hat p_{a,{\rm in}}(\omega)
\nonumber\\
&+\frac{G_{cs}\chi'_S(\omega)\sqrt{\kappa\gamma_{qE}}}
{g\,\chi_m(\omega)\,\sqrt{2\gamma_m\kappa}}\,
\Big[\tfrac{\chi_S(\omega)}{\chi'_S(\omega)}\big(\Delta_{qe}\chi'_S(\omega)+1\big)\hat x_{S,{\rm in}}(\omega)\\ \notag
&+\hat p_{S,{\rm in}}(\omega)\Big].
\end{align}
Examining the spectral density of the added force provides information about the sensitivity and overall performance of the system under the CQNC condition.
\section{Results and Discussions}
We now proceed to the numerical analysis of the power spectral density which provides a measure of the force sensitivity \cite{Motazedifard_2016}. The force sensitivity can be defined as,
% \begin{equation*}
    ${s}_F(\omega)
    = \sqrt{\hbar m_s \gamma_m}\, S_F(\omega),\
    $
  where, $S_F(\omega)$ quantifies the noise spectral density for an
equivalent input force, $F$.
% \end{equation*}
Thus, a lower value of $S_F(\omega)$
corresponds directly to an improved force sensitivity. 

Following Ref.~\cite{wimmer2014coherent}, the spectral density of the added noise for our setup (see \Cref{faddd}) is defined as follows.

 \begin{align}
 S^{add}_F(\omega)\delta(\omega-\omega^\prime)=\frac{1}{2} (\langle\hat F_{add}(\omega)\hat F_{add}(-\omega)\rangle +c.c.)
 \label{S_F}
\end{align} 

\begin{table}[h!]
\centering
\caption{Experimentally feasible parameters for the hybrid optomechanical system with a QD ensemble.}
\label{tab:params}
\begin{tabular}{lcc}
\toprule
\textbf{Parameters} & \textbf{Symbols} & \textbf{Typical Values} \\
\midrule
Cavity decay rate & $\kappa$ & $2\pi\times 1~\mathrm{MHz}$ \\
Mechanical frequency & $\Omega$ & $2\pi\times 10~\mathrm{MHz}$ \\
Mechanical damping rate & $\gamma_m$ & $2\pi\times 100~\mathrm{Hz}$ \\
QD-ensemble linewidth & $\gamma_{qE}$ & $2\pi\times 200~\mathrm{Hz}$ \\
Optomechanical coupling & $g_0$ & $2\pi\times 10~\mathrm{Hz}$ \\
Cavity--QD coupling & $G_{cs}$ & $2\pi\times 10~\mathrm{Hz}$ \\
OPA nonlinear gain & $\mathcal{G}$ & $0.3\,\kappa$ \\
OPA phase & $\theta$ & $0$--$\pi$ \\
Input laser wavelength & $\lambda_L$ & $1064~\mathrm{nm}$ \\
Input power & $P_L$ & $10^{-3}$–$10^6~\mathrm{pW}$ \\
Temperature & $T$ & $300~\mathrm{K}$ (room temperature) \\
\bottomrule
\end{tabular}
\end{table}
Using this we obtain the expression of spectral density of the added noise as
\begin{align}
   S_F^{add} =
   \notag
   & \frac{K_B T}{\hbar \Omega}  + \frac{1}{2} \frac{1}{g^2 |\chi_m(\omega)|^2(2\gamma_m \kappa)} \big|\frac{\lambda_-(\omega)\kappa - 1}{\lambda_-(\omega)}\big|^2 \\
   &+ \frac{1}{2}\frac{\omega ^2 +\Omega ^2 + \gamma_{qE} ^2/4}{\Omega^2}
   \label{SF_added_1}
\end{align}
In order to study the measurement added noise reduction, the Brownian thermal noise background can be avoided\cite{Motazedifard_2016}. By minimizing the spectral density, we obtain the minimum noise spectral density for our proposed scheme in hybrid optomechanical model as,
\begin{align}
    S_{F,CQNC}=\frac{1}{2}\frac{\omega^2+\Omega^2 +\gamma_{qE}^2/4}{\Omega^2}
\label{SF_CQNC}
\end{align}
We compare the efficiency of our hybrid system under CQNC condition by comparison with the standard optomechanical system.
For such standard optomechanical system, the noise spectral density can be written as \cite{RevModPhys.86.1391},
\begin{align}
  S_F(\omega)=  \frac{K_B T}{\hbar \Omega}+\frac{1}{2}\frac{\kappa}{\gamma_m} \frac{1}{g^2 |\chi|^2} \frac{1}{4}+ 4g^2 \frac{1}{\kappa\gamma_m}
\label{SF_SQL}
\end{align}
 The first term indicates the contribution of the thermal noise, second term represents shot noise and the third term corresponds to the back-action noise. Minimizing this with respect to the laser power, we can get the SQL value,
\begin{align}
    S_{F,SQL}=\frac{1}{\gamma_m |\chi_m|}
\label{SQL_min}
\end{align}
Here, the thermal force is considered to be negligibly small. The dip of the noise spectral density occurs at an optomechanical coupling strength, $ g_{SQL} = \sqrt{\kappa}/(2\sqrt{| \chi _m |})  
   \label{g_{SQL}} $. 
\begin{figure}[h!]
\centering
    \includegraphics[height=5cm,clip]{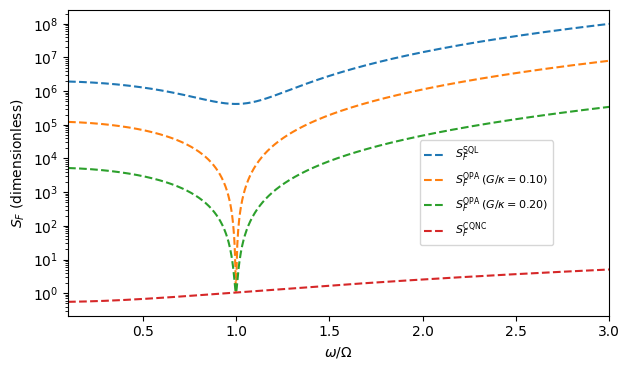}
\caption{\justifying Noise Power Spectral Density for standard optomechanical system, optomechanical system with OPA for parametric gain $G/\kappa = 0, 0.1$ and $0.3$ and optomechanical hybrid system with CQNC scheme. The spectral densities are normalized by $\hbar m \omega _m \gamma _m$ in order to be represented in units of $\rm{N^2 Hz^{-1}}$. The different lines in the plot represent the standard optomechanical system (blue curve), the hybrid electro-optomechanical system with OPA (orange and green curves), and the hybrid system with the CQNC scheme (the red line at the bottom). The parameters used \cite{wimmer2014coherent, singh2023enhanced}: $g_0 = 300 \times 2\pi$ Hz, $\Omega= 300 \times 2\pi$ KHz, $\gamma_m = 30 \times 2\pi$ Hz, $\kappa= 2\pi$ MHz, $P=100$ mW, $\omega_L = 384 \times 2\pi$ THz}.

\label{PSD1}
\end{figure}

\begin{figure}[h!]
\centering
    \includegraphics[height=5cm,clip]{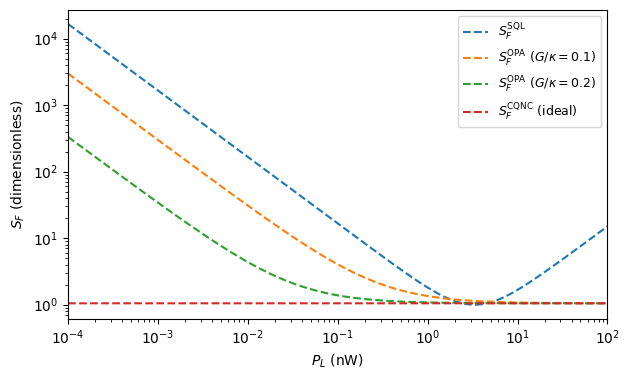}
    \caption{\justifying Noise Power Spectral Density at resonance (i.e., $\omega=\Omega$) as a function of laser driving power for the standard optomechanical system and the electro-optomechanical hybrid system with the CQNC scheme. The blue line represents the standard optomechanical system, whereas, the orange and green curve indicate hybrid system with OPA gain $G/\kappa = 0.1$ and $0.3$ respectively. (Parameters are the same as in Fig.\ref{PSD1})}
\label{PSD_PL}
\end{figure}

% In Fig.~\ref{PSD1}, we show the noise spectral density of the optomechanical system as a function of the detection frequency. When the system is driven at resonance, i.e.\ when the detection frequency is $\Omega$, the spectrum $S_F$ exhibits a clear dip. At this point, the values of $S_F$ for the latter two parameter sets are nearly identical. 

% However, when the OPA gain $G$ is increased from $0.1$ to $0.3$, the noise spectral density at off-resonant frequencies is noticeably reduced compared to the standard optomechanical case, as can be seen from the corresponding curves. In the CQNC configuration, we further observe that $S_F$ is suppressed by several orders of magnitude at frequencies both below and above the mechanical resonance. Although CQNC does not significantly enhance the sensitivity exactly at resonance compared to standard system, it leads to a substantial reduction of the noise at off-resonant frequencies, thereby improving the broadband force-sensing performance.
We first evaluate the noise spectral density at the mechanical resonance, $\omega=\Omega$, where a pronounced dip is observed (Fig.~\ref{PSD1}). At resonance, the hybrid system yields nearly the same sensitivity for both OPA gain values, whether CQNC condition is absent or included. By contrast, the behavior changes remarkably away from resonance. Increasing the OPA gain from $G = 0.1$ to $G = 0.3$ leads to a clear reduction of the noise spectral density at off-resonant frequencies when compared with the standard optomechanical system, as reflected in the corresponding curves. In the presence of CQNC, the suppression becomes even more prominent and $S_F$ is reduced by several orders of magnitude for frequencies both below and above $\omega=\Omega$.
Thus it provides a strong noise reduction over a broad range of off-resonant frequencies beyond SQL.

% The variation of noise PSDs vs laser driving power is shown in Fig. [\ref{PSD_PL}]. The laser driving power is proportional to the square of optomechanical coupling given by $P_L = 2\hbar {\omega _L} \kappa (g/g_0)^2$. The noise spectral density is expressed as a function of input laser power. 
% The figure shows that in a standard optomechanical system (represented by the blue line), the noise spectral density initially decreases as laser power increases. However, after reaching the minimum point, it begins to rise again due to the significant back-action noise at higher laser power levels. Whereas, for the hybrid system (orange and green line corresponding to OPA gain 0.1 and 0.3 respectively) the PSD decreases with the laser power and after reaching the minimum, it does not increase any more. This happens because, when CQNC scheme is incorporated, it cancels out the back-action noise at higher laser driving power. 

We next examine how the noise power spectral density (PSD) depends on the input laser power, as shown in Fig.~\ref{PSD_PL}. The laser power is related to the optomechanical coupling via
$
    P_{L} = 2\hbar \omega_{L}\kappa \left(\frac{g}{g_{0}}\right)^{2},
$
so that the noise spectral density can be viewed directly as a function of $P_{L}$. For the standard optomechanical system (blue curve), the PSD initially decreases with increasing laser power, reaches an optimum minimum, and then increases again. In contrast, for the hybrid system incorporating the OPA (orange and green curves, corresponding to OPA gains $G = 0.1$ and $G = 0.3$, respectively), the PSD decreases with increasing laser power and, after attaining its minimum value, does not increase again. This behavior is a direct consequence of the CQNC mechanism, which cancels the back-action contribution even at relatively high driving powers and thereby prevents the PSD from rising.

Moreover, increasing the OPA gain from $G = 0.1$ to $G = 0.3$, shifts the minimum of the PSD to a lower value of $P_{L}$. Thus, within our framework, the use of an OPA with higher gain allows one to reach the minimal spectral density at reduced input power.

\section{Imperfect Matching}
Ideally, perfect back-action noise cancellation requires accurate matching of both the coupling strengths and the susceptibilities of the relevant modes, as discussed in Sec.~III. In practice, however, achieving exact equality of the decay rates and therefore of the susceptibilities is extremely challenging, and the coupling strengths can only be tuned with finite precision. Since some degree of mismatch in both couplings and decay rates is unavoidable, it is essential to investigate the performance of the system under imperfect CQNC conditions as well.

\textbf{(i) $\gamma_{qE} \neq \gamma_m$}\\
We first consider the situation in which all CQNC conditions are satisfied except for the matching of the decay rates. A small relative mismatch $\delta$ between the decay rates of the QD ensemble and the mechanical mode can be introduced as
\begin{equation}
    \delta = \frac{\gamma_{qE} - \gamma_m}{\gamma_m},
\end{equation}
so that $\gamma_{qE} \neq \gamma_m$ when $\delta \neq 0$. In the presence of this mismatch, the noise spectral density modifies to
\begin{align}
    S_{F,\mathrm{CQNC}}
    = \frac{1}{2}\,\frac{\omega^2 + \Omega^2 + \gamma_{qE}^2/4}{\Omega^2}
      + \frac{\lambda_+^2 \kappa g^2}{\gamma_m}
        \left|1 + \frac{\chi'_S}{\chi_m}\right|^2.
    \label{CQNC_mismatched_gamma}
\end{align}
In this expression, the first term typically dominates over the second. The corresponding noise spectral densities are plotted in Fig.~\ref{Imperfect_gamma1}. As seen there, the curves for perfect CQNC (red line) and for the decay-rate–mismatched case (black dashed curve) are almost indistinguishable. Thus, even for a moderate mismatch of the decay rates (here, $\delta = 0.3$), the spectral density remains very close to the ideal CQNC limit.

%, which means even with a slight deviation in the decay rate matching condition, we can still get the advantage compared to SQL.

%Furthermore, from Fig. we see the variation of noise spectral densities as a function of laser power. The blue line represents SQL. The spectral densities decrease with increasing laser power. After hitting the minimum, the spectral density for SQL increases again, while the spectral densities for hybrid model for perfect matching of decay rates (orange and green line with OPA gain 0.1 and 0.3 respectively) does not increase any more due to complete cancellation of back-action noise. For the imperfect case ($\delta = 0.3$), the spectral densities increases after touching  the minima, with increasing laser power. However, it still lie below the SQL.

%Therefore from the Fig. and Fig. it is evident that even with slightly unmatched decay rates ($\delta = 0.3$), noise reduction beyond SQL is achievable.
\begin{figure}[!ht]
\centering    

\includegraphics[width=.5\textwidth]{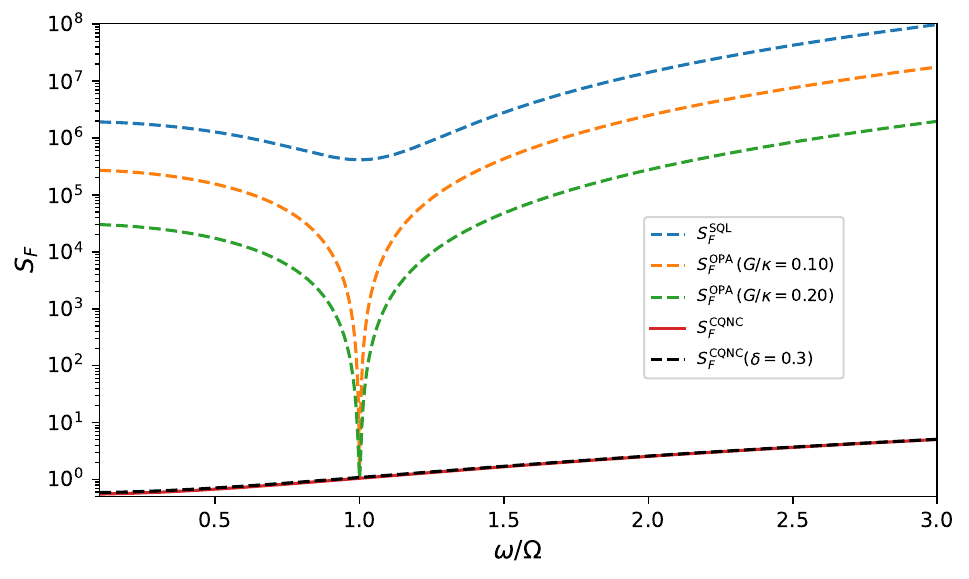}
      \caption{ \justifying Variation of noise spectral density with frequency. The blue solid line represents SQL, the orange and green dashed lines indicates the noise spectral densities  with the hybrid model with OPA gain 0.1 and 0.2 respectively. The red line is for perfect CQNC and the black dashed line represents the mismatched CQNC with $\delta = 0.3$. %coupled to an optomechanical system .
      \label{Imperfect_gamma1}}
\end{figure}

\textbf{(ii) $G^{\prime} \neq g$}\\
We now consider the case in which all CQNC conditions are satisfied except for the matching of the coupling strengths. A small relative mismatch $\epsilon$ between the effective coupling $G^{\prime}$ and the optomechanical coupling $g$ is defined as
\begin{equation}
    \epsilon = \frac{G^{\prime} - g}{g},
\end{equation}
so that $G^{\prime} \neq g$ when $\epsilon \neq 0$. In the presence of this coupling mismatch, the noise spectral density takes the form %acquires a residual back-action contribution 
\begin{align}
    S_{F,\mathrm{CQNC}}
    = \frac{1}{2}\,\frac{\omega^2 + \Omega^2 + \gamma_{qE}^2/4}{\Omega^2}
      + \frac{\lambda_+^2 \kappa g^2}{\gamma_m}
        \left| 1 - \frac{G^{\prime 2}}{g^2} \right|^2.
    \label{CQNC_mismatched_G}
\end{align}

The corresponding spectra are shown in Fig.~\ref{Imperfect_G}. As illustrated there, the noise spectral density for perfect CQNC (purple curve) lies below that of the coupling-mismatched case (red curve), reflecting the presence of a small residual back-action term. Nonetheless, even under this imperfect condition, the noise level still surpasses the SQL by approximately five orders of magnitude.

Therefore from the Fig. \ref{Imperfect_gamma1} and Fig. \ref{Imperfect_G} it is evident that even with small deviation from the perfect CQNC conditions ($G^{\prime} = g$ and $\gamma_{qE} = \gamma_m$), the hybrid system ensures the possibility of maintaining the advantage of CQNC scheme in hybrid electro-optomechanical system.
%Moreover, Fig. shows variation of the noise spectral density with laser power. Here also we see that with unmatched coupling strength ($\epsilon = 0.1$), the spectral densities for the hybrid system with OPA gain 0.1 and 0.3 lies below the SQL. Therefore from the Fig. 3 and Fig. 4 it is evident that even with slightly unmatched coupling strengths, noise reduction beyond SQL is possible.
\begin{figure}[!ht]
\centering
      \includegraphics[width=.5\textwidth]
      {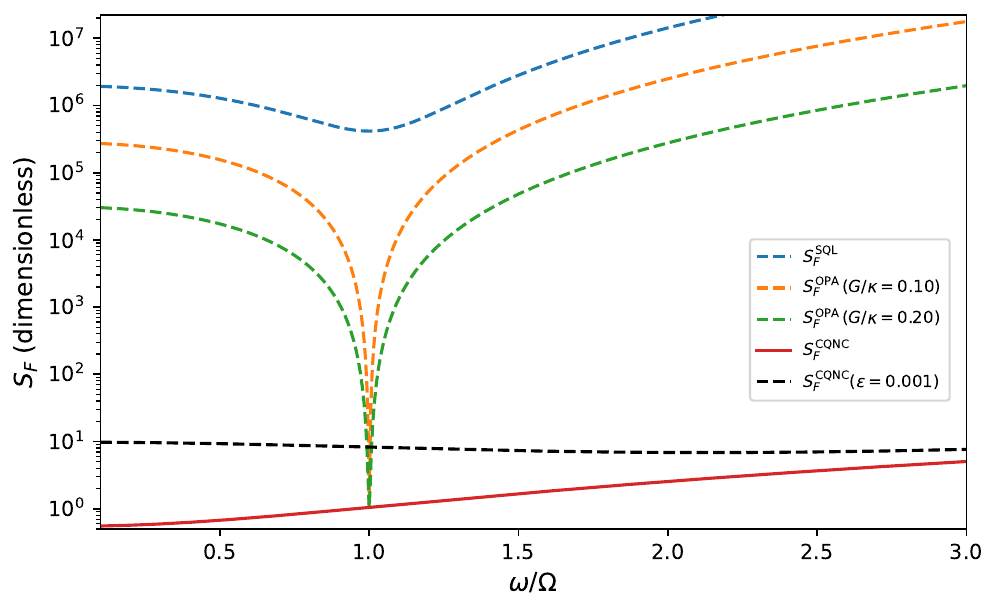}
      \caption{\justifying Variation of noise spectral density with frequency. The blue solid line represents SQL, the orange and green dashed lines indicates the noise spectral densities  with the hybrid model with OPA gain 0.1 and 0.2 respectively. The red line is for perfect CQNC and the black  line represents the mismatched CQNC with $\epsilon = 0.01$. %coupled to an optomechanical system .
      \label{Imperfect_G}}
\end{figure}
\section{Conclusion}
We have proposed and analyzed a hybrid cavity--optomechanical platform in which an ensemble of quantum dots (QDs) is coupled to the intracavity optical mode, while a optical parametric amplifier (OPA) modifies the optical quadratures. Working in the linearized regime, we derived the corresponding quantum Langevin equations and identified compact frequency-domain conditions that make explicit how the QD-induced susceptibility modifies the optical response. These conditions provide a natural route to implement coherent quantum-noise cancellation (CQNC) in a hybrid optomechanical system

Within this CQNC framework, we studied the spectral density of the added force noise as a measure of the force-sensing sensitivity. First, by examining the dependence of the spectral density on the detection frequency, we showed that the incorporation of the CQNC scheme allows the sensitivity to surpass the standard quantum limit (SQL) by up to six orders of magnitude over a broad frequency range. We then investigated the spectral density as a function of the input laser power $P_L$. Our analysis reveals that, as the OPA gain is increased, the minimum of the spectral density is reached at progressively lower laser powers, indicating that the device can operate in a low-power regime while still achieving optimal noise suppression.

From an experimental perspective, realizing ideal CQNC is challenging, since it requires precise matching of both the coupling strengths and the decay (or dephasing) rates of the relevant modes. We therefore examined the impact of imperfect parameter matching. In particular, we considered (i) a mismatch between the optomechanical coupling and the effective cavity--QD-ensemble coupling, and (ii) unequal mechanical damping and QD-ensemble dephasing rates. In both cases, we find that moderate deviations from the ideal matching conditions still yield a substantial reduction of the added-noise spectral density compared to a standard optomechanical system without CQNC, and the SQL can remain surpassed by several orders of magnitude.

These results demonstrate that a QD-ensemble-based hybrid optomechanical platform, assisted by an OPA, can provide a robust and tunable route to eliminate back-action noise in force detection. Looking at the potentially integrable nature, such a system could be of significant interest for applications in quantum information processing and communication, gravitational-wave and precision force detection, and more broadly in quantum metrology and noise-engineered quantum technologies.

\section*{Acknowledgement}
A.R. gratefully acknowledges the support of a research fellowship
from UGC, Government of India. A.K.S. acknowledges the
grant from MoE, Government of India (Grant No. MoESTARS/STARS-2/2023-0161).
\section*{\large Appendix}
\appendix

\section{ Appendix : Collective description of the QD ensemble}
\label{app:QDensemble}

We consider an ensemble of $N$ identical two-level quantum dots (QDs), each with ground state $\lvert g_i \rangle$ and excited state $\lvert e_i \rangle$, coupled to a single optical cavity mode $\hat{a}$. The lowering and raising operators of the $i$th QD are defined as
\begin{equation}
    \sigma_i^{-} = \lvert g_i \rangle \langle e_i \rvert, 
    \qquad
    \sigma_i^{+} = \lvert e_i \rangle \langle g_i \rvert.
\end{equation}
In the rotating-wave approximation, the Hamiltonian describing the ensemble and its coupling to the cavity field can be written as
\begin{equation}
    H_{\mathrm{QD}} 
    = \hbar \sum_{i=1}^{N} \omega_{qe}\, \sigma_i^{+} \sigma_i^{-}
      + \hbar g_{0}(\hat{a}+\hat{a}^\dagger) \sum_{i=1}^{N}  \bigl(  \sigma_i^{-} + \,\sigma_i^{+} \bigr),
    \label{eq:H_QD_sigma}
\end{equation}
where $\omega_{qe}$ is the transition frequency of the QDs and $g_{0}$ is the single-QD coupling strength to the cavity mode.

It is convenient to introduce collective spin operators for the ensemble using Holstein-Primkoff [HP]~\cite{PhysRev.58.1098} transformation,
\begin{equation}
    \hat{S}^{-} = \frac{1}{\sqrt{N}} \sum_{j=1}^{N} \sigma_j^{-},
    \qquad
    \hat{S}^{+} = \frac{1}{\sqrt{N}} \sum_{j=1}^{N} \sigma_j^{+},
    \label{eq:collective_S}
\end{equation}
which describe the symmetric collective excitation of the QDs. In terms of these operators, the Hamiltonian in Eq.~\eqref{eq:H_QD_sigma} can be rewritten as
\begin{equation}
    H_{\mathrm{QD}}
    = \hbar \omega_{qe}\, \hat{S}^{+} \hat{S}^{-}
      + \hbar G_{cs} (\hat{a}+\hat{a}^\dagger) \bigl(  \hat{S}^{-} + \,\hat{S}^{+} \bigr),
    \label{eq:H_QD_collective}
\end{equation}
% where the collective light--matter coupling is enhanced to
% \begin{equation}
%     G = g_{0} \sqrt{N}.
% \end{equation}
\vspace{10mm}
\section{Appendix: Derivation of the cavity phase quadrature $P_a(\omega)$}
\label{appendixB}
Starting from the linearized quantum Langevin equations in the frequency domain, the
quadrature variables satisfy the following set of algebraic relations
\begin{align}
 X_a &= A_1 X_a^{\mathrm{in}},\\
P_a &=  A_2 X_a^{\mathrm{in}} - B_2 F(\omega) - C_2 X_S + D_2 P_a^{\mathrm{in}},\\
X_b &= - A_3 X_a^{\mathrm{in}} + B_3 F(\omega),\\
P_b&=  A_4 X_b,\\
X_S &=- A_5 P_S + B_5 X_S^{\mathrm{in}},\\
P_S& =- A_6 X_S^{\mathrm{in}} - B_6 X_a^{\mathrm{in}} + D_6 P_S^{\mathrm{in}},
\end{align}
where \(F(\omega) = F_{\mathrm{th}}(\omega) + F_{\mathrm{ext}}(\omega)\) denotes the sum of thermal and external forces, and the coefficients \(A_i,B_i,C_2,D_i\) are defined in terms of the system parameters as
\[
A_1=\sqrt{\kappa_a}\lambda_+,\;
A_2=g^{2}\chi_m\lambda_+\lambda_-\sqrt{\kappa_a},\;
B_2=\chi_m\sqrt{\gamma_m}\,g\lambda_-,\; \] 
\[
C_2=G_{\mathrm{OM}}\lambda_-,\;
D_2=\sqrt{\kappa_a}\lambda_-,
A_3=g\chi_m\sqrt{\kappa_a}\lambda_+,\;
\]
\[
B_3=\chi_m\sqrt{\gamma_m},\;
A_4=\frac{i\omega}{\omega_m},\;
A_5=\chi_m\omega_m,\;
B_5=\chi_m\sqrt{\kappa_M},
\]
\[
A_6=\chi_m' \sqrt{\gamma_m},\;
B_6=G_{\mathrm{OM}}\xi\lambda_+\sqrt{\kappa_a},\;
D_6=\xi\sqrt{\gamma_{qE}}.
\]

We obtain a closed expression for the cavity phase quadrature in terms of the input noises and the total force:
\begin{align}
P_a(\omega) &= \bigl(A_2 + A_5 B_6 C_2 \bigr) X_a^{\mathrm{in}}(\omega)
      + D_2 P_a^{\mathrm{in}}(\omega)
      + A_5 A_6 C_2 X_S^{\mathrm{in}}(\omega) \notag\\
      &\quad - A_5 C_2 D_6 P_S^{\mathrm{in}}(\omega)
      - B_5 C_2 X_S^{\mathrm{in}}(\omega)
      - B_2 F(\omega).
\label{eq:Pa_omega_appendix}
\end{align}
Equation~\eqref{eq:Pa_omega_appendix} is the starting point for evaluating the added-force noise spectral density and the corresponding force-sensing sensitivity discussed in the main text.

\bibliography{bibb}
\end{document}